\documentclass[aps,floatfix,superscriptaddress,notitlepage,showkeys]{revtex4}
\usepackage{amsmath,amssymb,amsfonts,graphics,graphicx,dcolumn,bm,enumerate}
\usepackage{comment,natbib}
\usepackage{multirow,color}

\newcommand{\aalto}
{\affiliation{Department of Computer Science, Aalto University School of 
Science,
P.O. Box 15400, FI-00076 AALTO, Finland}}
\newcommand{\cmp}
{\affiliation{Condensed Matter Physics Division, 
Saha Institute of Nuclear Physics, 1/AF Bidhannagar, Kolkata 700064, India.}}
\newcommand{\isi}
{\affiliation{Economic Research Unit, Indian Statistical Institute, 203 B. T. Road, Kolkata 700108, India.}}
\newcommand{\hok}
{\affiliation{Graduate School of Information Science \& Technology,  Hokkaido University, N14-W9, Kita-ku, Sapporo 060-0814, Japan.}}

\begin{document}
\title{Inequality measures in kinetic exchange models of wealth distributions}

\author{Asim Ghosh}
\email[Email: ]{asim.ghosh@saha.ac.in}
\cmp \aalto
\author{Arnab Chatterjee}%
\email[Email: ]{arnabchat@gmail.com}
\cmp
\author{Jun-ichi Inoue}
\thanks{deceased}
\hok
\author{Bikas K. Chakrabarti}%
\email[Email: ]{bikask.chakrabarti@saha.ac.in}
\cmp \isi

\begin{abstract}
In this paper, we study the inequality indices for some models of wealth 
exchange. We calculated Gini index and newly introduced $k$-index and 
compare the results with reported empirical data available for different 
countries. We have found lower and upper  bounds for  the indices and discuss 
the efficiencies of the models. Some exact analytical calculations are given 
for 
a few cases. We also exactly compute the quantities for Gamma and double Gamma 
distributions.
\end{abstract}

\maketitle

\section{Introduction}

Socio-economic  
inequality~\cite{arrow2000meritocracy,stiglitz2012price,neckerman2004social,
goldthorpe2010analysing} 
is manifested in the existence of unequal rewards and opportunities  for social
positions or statuses in a society. Structured, recurrent
patterns of unequal distributions of goods, wealth, opportunities,
rewards and punishments are mainly measured in terms of  
\textit{inequality of conditions}, and \textit{inequality of opportunities}.
The former refers to the unequal distribution of income,
wealth and material goods, while the latter
refers to the unequal distribution of `life chances' of individuals. 
This is somehow reflected in measures such as level of education, health 
status, and treatment by the criminal justice system.
Socio-economic inequality often results in crisis, political unrest and 
instability, 
conflict, war, criminal activity and finally affects economic 
growth~\cite{hurst1995social}.
Initially, economic inequalities were studied in the context of income and  
wealth~\cite{Yakovenko:RMP,chakrabarti2013econophysics,aoyama2010econophysics},
but the notions and observations have led to widespread research, see e.g. 
Ref.~\cite{chatterjee2014socio,chatterjee2015social} for various socio-economic 
inequalities.
The study of inequality in society~\cite{Cho23052014,Chin23052014,Xie23052014}
is a topic of global focus and utmost current interest, bringing together
researchers from various disciplines.

By the end of the 19th century, 
Pareto~\cite{Pareto-book} made extensive studies and found that 
wealth distribution in Europe follows a power law 
for the rich, commonly known to be the \textit{Pareto law}.
Subsequent studies have revealed that the distributions of income and wealth 
possess some globally
robust features (see, e.g., \cite{chakrabarti2013econophysics}):
the bulk of both the income and wealth distributions seem to reasonably fit 
both the log-normal 
and the Gamma distributions. Economists have a preference for the log-normal 
distribution~\cite{Montroll:1982,gini1921measurement}, while 
statisticians~\cite{Hogg-2007} and 
physicists~\cite{Chatterjee:EWD,Chatterjee2007,Yakovenko:RMP} 
root for the Gamma distribution for the probability density or 
Gibbs/exponential 
distribution for the corresponding cumulative distribution. 
The high end of the distribution, known as the `tail', is well described by a 
power law as observed by Pareto. Formally,
the probability distribution of wealth is given by
\begin{equation}
\label{par}
P(m) \sim
\left\{ \begin{array}{lc}
F(m) & \textrm{for} \ m < m_c,\\
\frac{\alpha m_c^\nu}{ m^{1+\nu}} \ \ \ \ & \textrm{for} \  m \ge m_c,
\end{array}\right.
\end{equation}
 where $\alpha$ is a constant and 
 $\nu$ is called the Pareto exponent, ranging between 1 and 
3~\cite{chakrabarti2013econophysics} 
 (See Ref.~\cite{Wileybook2} for a historical account of Pareto's data and some 
recent sources). 
$F(m)$ is some function which could be exponential, Gamma or lognormal.
The crossover point $m_c$ is extracted from the numerical fittings. 

One of the key class of models uses the kinetic theory of 
gases~\cite{Dragulescu:2000},
where the gas molecules colliding and exchanging energy was mapped to agents
meeting to exchange wealth, following certain rules~\cite{Chatterjee2007}.
In these models, a pair of agents agree to trade, each save a fraction 
$\lambda$ of their 
instantaneous money/wealth and  exchanges a random fraction of the rest at each 
trading step.
The distribution of wealth in the steady state, $P(m)$ matches well with the 
empirical data.
When the saving fraction $\lambda$ is fixed, i.e., in case of homogeneous 
agents (CC model hereafter)~\cite{Chakraborti:2000}, 
$P(m)$ are very well approximated to Gamma distributions~\cite{Patriarca2004}.
It is important to note that, in reality,
the richest follow a different dynamic where heterogeneity plays the key role. 
To obtain the power law distribution of wealth for the richest, one needs 
simply to consider
each agent as different in terms of the fraction of wealth he/she saves in 
each trading~\cite{Chatterjee:2004},
which is very natural to assume, because it is quite likely that agents in a 
market think differently
from one another.
With this very little modification, one can explain the whole range of wealth 
distribution~\cite{Chatterjee2007}.
When $\lambda$ is distributed uniformly in $[0,1)$ and quenched, (CCM 
model hereafter),  i.e., for heterogeneous agents, one obtains a Pareto law for 
the probability density of wealth
$P(m) \sim m^{-\nu}$ with exponent 
$\nu=2$~\cite{Chatterjee:2004,Chatterjee2007}.
Several variants of these models, find possible applications
in a variety of trading 
processes~\cite{chakrabarti2013econophysics,pareschi2013interacting}.

Socio-economic inequalities are quantified in various ways. The most popular 
measures are absolute,
in terms of indices, e.g., Gini~\cite{gini1921measurement}, 
Theil~\cite{theil1967economics}, Pietra~\cite{eliazar2010measuring} 
and the recently introduced $k$ index~\cite{ghosh2014inequality}.
The alternative approach is a relative measure,
in terms of probability distributions of various quantities, but the most of 
the above mentioned indices
can be computed from the distributions. Most quantities often 
display broad distributions, usually lognormals, power-laws or their 
combinations.
For example, the distribution of income is usually an exponential followed by a 
power 
law~\cite{druagulescu2001exponential} (see 
Ref.\cite{chakrabarti2013econophysics} for other examples).

\begin{figure*}[t]
\includegraphics[width=5.5cm]{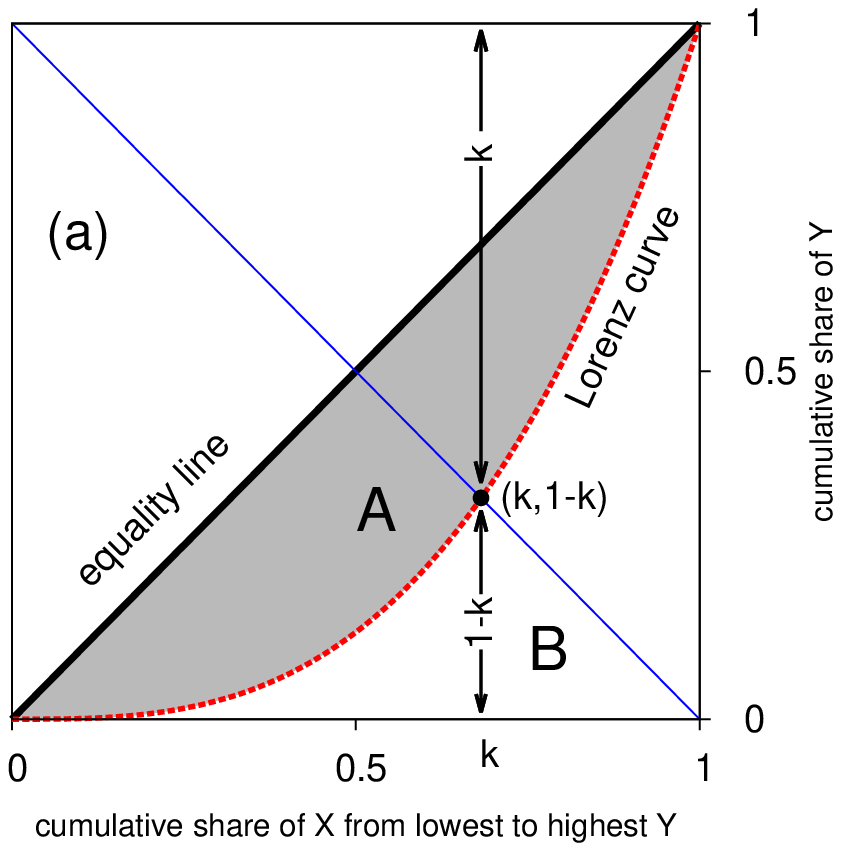}
\hskip -0.5cm
\includegraphics[width=12.5cm]{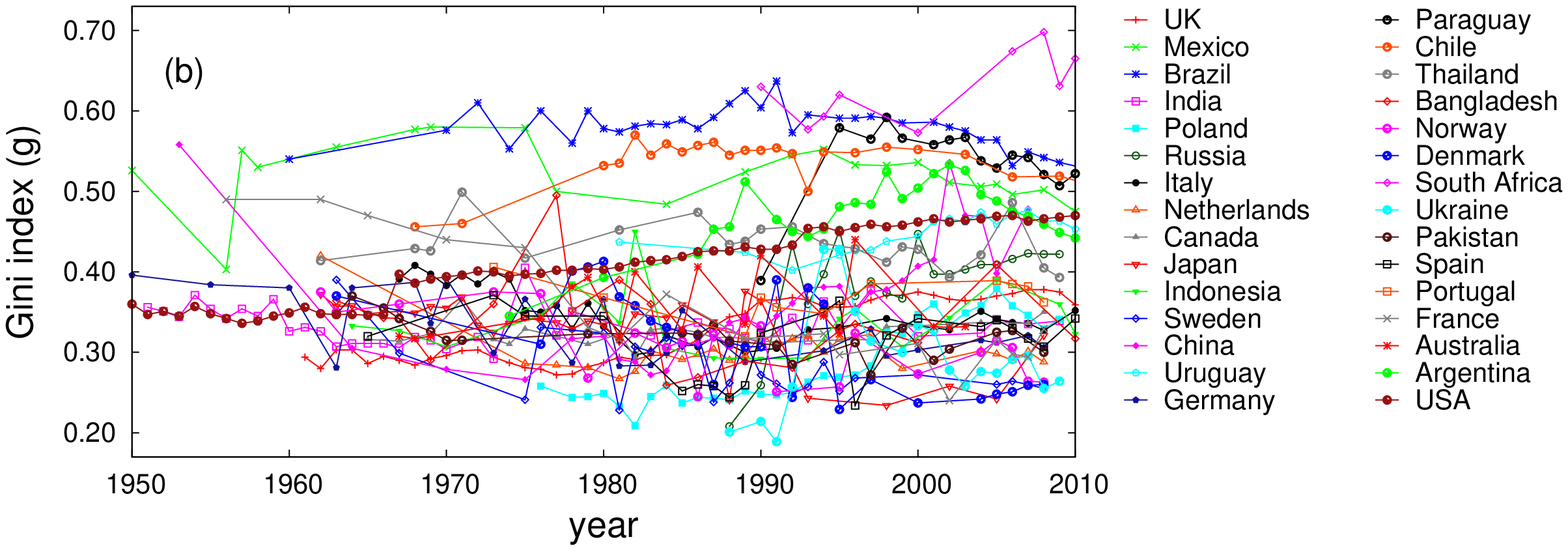}
\caption{(a) Gini and $k$-index schematic: the solid red line is the Lorez 
curve, the cumulative proportion $X$ of ordered individuals (from lowest to 
highest) in terms of the cumulative proportion of their sizes $Y$.
Gini index is given by the ratio  $g=A/(A+B)$.
The $k$-index is computed from the intersection ($k,1-k$) of the reverse 
diagonal with the Lorenz curve.
(b) Gini index from World Bank data~\cite{worldbank} for several countries 
over years.
}
\label{fig:emp}
\end{figure*}

To compute the Gini index, one has to 
consider the Lorenz curve~\cite{Lorenz}, that represents 
the cumulative proportion $X$ of ordered individuals (from lowest to highest) 
in terms of the cumulative proportion of their sizes $Y$ (See 
Fig.~\ref{fig:emp}a).
$X$ can represent income or wealth of individuals.
The Gini index ($g$) is defined as the ratio between the area 
enclosed between the Lorenz curve and the equality line, to that below the 
equality line.
If the area between 
(i) the equality line and  the Lorenz curve is $A$, and 
(ii) that below the Lorenz curve as $B$,
the Gini index is given by $g=A/(A+B)$.
The recently introduced `$k$ index'~\cite{ghosh2014inequality} is 
defined as the fraction $k$ such that  $(1-k)$ fraction 
of individuals possess $k$ fraction of income or 
wealth (See Fig.~\ref{fig:emp}a)~\cite{inoue2015measuring}.

In this paper, we investigate the inequality in wealth in some models of 
wealth distribution which are inspired by kinetic theory of gases. 
We mainly discuss the results for two well studied models (CC and CCM) 
and a new model for bimodal distribution of wealth.
We numerically compute the inequality indices, Gini index and $k$-index to 
quantify the inequalities. 
Gini index $g$, the most popular and widely used measure for inequality 
in case of income and wealth distribution, can take value from $0$ to $1$. The 
value $g=0$ refers to complete equality and $g=1$  represents completely 
inequality. The meaning of $k$-index, say, for a wealth distribution, is the 
following:
$k$ fraction of the top wealthiest people possess $1-k$ fraction of total 
wealth.
We found that in both CC and CCM models there are some upper and lower limits 
of the indices. 
For CC model, $g$ varies between $0$ and $0.5$, and $k$ from $0.5$ to $0.68$. 
Similarly, for CCM model, $g$ varies between $0.4$ to $0.85$. 
Therefore, both models independently do not cover the possible theoretical 
range of values of $g$ and $k$. 
We find that the range of $g$ as found from empirical data ($0.2-0.7$)  (see 
Fig.~\ref{fig:emp}, using World Bank data~\cite{worldbank}) can be 
well covered by CCM model.
We also considered a model where two groups of agents have fixed but different 
saving propensities. Depending on the combinations, the resulting probability 
distribution of wealth is found to be unimodal or bimodal. The phase boundaries,
depending on the ratio of the two groups and the combination of values of 
their saving propensities are also computed numerically.
The bimodal distribution seems to fit well to a combination of two gamma 
distributions (double -Gamma distribution).
Gini index and $k$ index are calculated for this model for different 
combination of parameters.
Next, we considered gamma and double-Gamma distribution and computed certain 
quantities like Lorenz curve and Gini indices. 

\section{Models and numerical simulation results}
\label{sec:model}
Kinetic exchange models of wealth distributions~\cite{Chatterjee2007} serve as
simple paradigmatic models for exchange of wealth in an economy. The main idea 
is that
agents possess wealth $m_i$ which is redistributed upon trading with others.
The `economy' is assumed to be a `closed' one, in the sense that neither the 
number of agents change nor does that total amount of wealth in the system, and 
the economic activity is limited to exchange of wealth according to certain 
rules.
The basic model in the framework is just the random sharing of wealth,
motivated by random exchange of energy between gas molecules, as in the 
framework
of kinetic theory of gases~\cite{Yakovenko:RMP}.
The basic money exchange model~\cite{Dragulescu:2000}
imitates the 
kinetic exchange in an ideal gas, but subsequently developed models 
incorporate the notion of `savings'. In the following,
we discuss these models, and also compute the inequality measures like
Gini and $k$-index.

\subsection{CC model}
\begin{figure}[h]
\includegraphics[width=8.9cm]{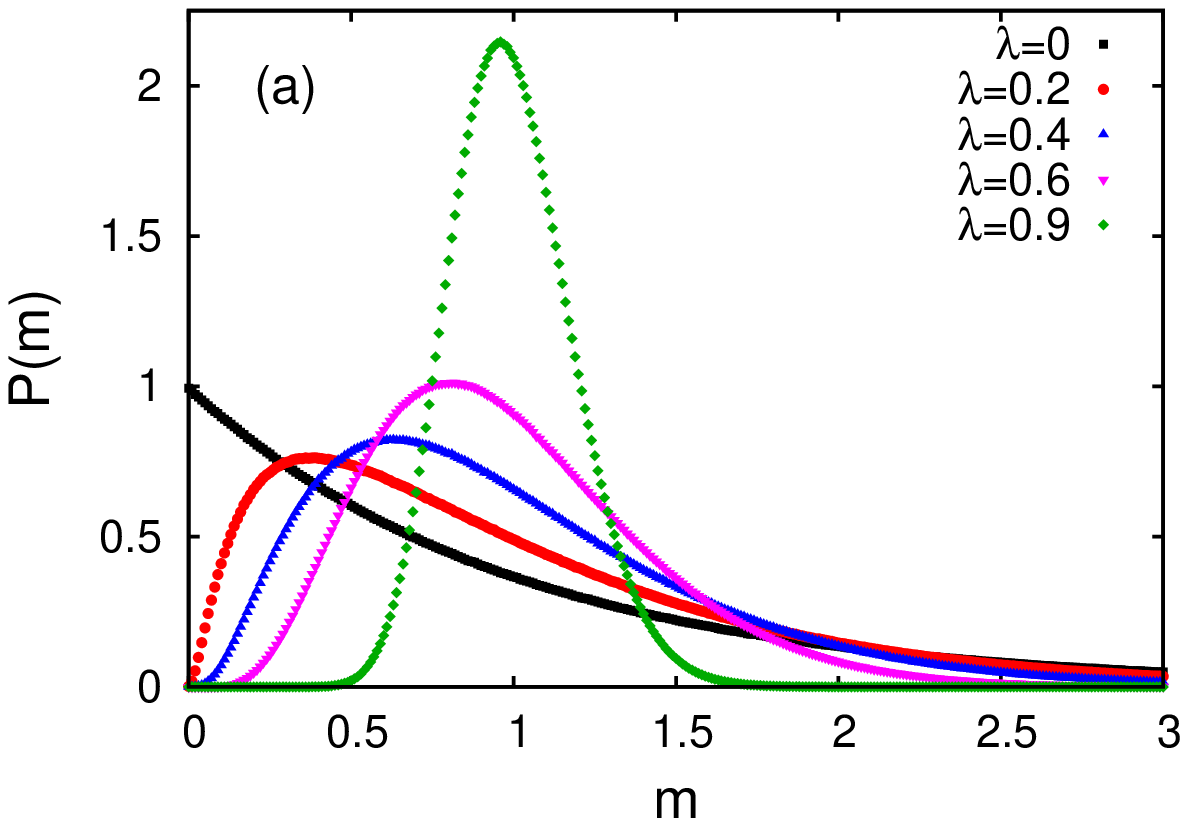}
\includegraphics[width=8.9cm]{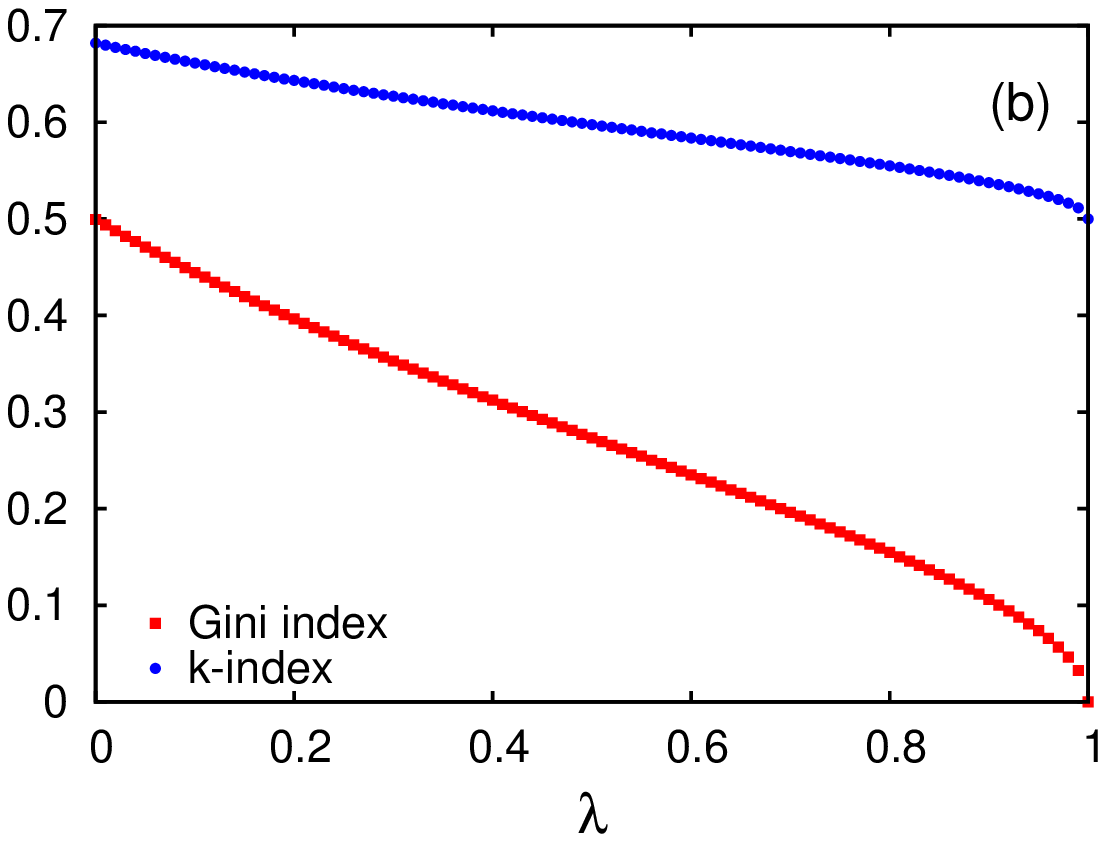}
\caption{(a) Probability distributions $P(m)$ for different values of 
$\lambda$ in the CC model.
(b) Gini and $k$-index for the entire range of $\lambda$ in CC model.
Simulation results are shown for $N=1000$ agents with average wealth unity.
}
\label{fig:cc}
\end{figure}

Savings come as a natural ingredient in a trading economy.
In each trading step, a pair of agents exchange their wealth 
in the following way:
they keep a fixed fraction $\lambda$ of their wealth to themselves
and the rest $1-\lambda$ fraction is pooled up to be randomly split among the 
two~\cite{Chakraborti:2000}.
In this model (CC model hereafter), agents are homogeneous -- all of them save 
the same fraction of their instantaneous wealth
at each trading step.
Formally, the dynamics is defined by
\begin{equation}
\begin{aligned}
m_i(t+1) &= \lambda m_i(t)+r(1-\lambda)\left (m_i(t)+m_j(t)\right) \\
m_j(t+1) &= \lambda m_j(t)+(1-r)(1-\lambda)(m_i(t)+m_j(t)),
\end{aligned}
\label{eq:cc}
\end{equation}
where $r$ is a random fraction in $[0,1]$, drawn in each time (exchange) step. 
$m_i(t)$ and $m_i(t+1)$ are the wealth of the $i$th agent at trading times $t$ 
and $(t+1)$
respectively.
The `saving propensity' $\lambda$ is a fixed fraction in $[0,1)$. $\lambda=0$ 
corresponds to complete
random exchange (DY model) while $\lambda=1$ gives no dynamics.
For $\lambda=0$, $P(m)=\exp(-m/\langle m \rangle)$ is exponential,
for which Lorenz curve, Gini and $k$ index were 
derived~\cite{inoue2015measuring}.
However, for finite $\lambda$,
$P(m)=C m^\alpha \exp(-m/T)$, has a form of  Gamma 
distribution~\cite{Patriarca2004}, 
where $T=\frac{1}{1+\alpha}$ and 
$C=\frac{(\alpha+1)^{(\alpha+1)}}{\Gamma(\alpha+1)}$. 
The exponent $\alpha$ is related to the parameter $\lambda$ as
$\alpha = \frac{3\lambda}{1-\lambda}$.
We plot $P(m)$ vs. $m$ for different values of $\lambda$ in 
Fig~\ref{fig:cc}a.
For these simulations (and for each case discussed in the paper), the average 
wealth $\langle m \rangle$ is set to unity.
We measured inequality in the distributions in terms of Gini and $k$ index and 
plotted in Fig.~\ref{fig:cc}b for different values of $\lambda$.

\subsection{CCM model}
In this model~\cite{Chatterjee:2004,Chatterjee2007} (CCM model hereafter), 
agents are assumed to be heterogeneous, the saving fraction $\lambda$
for each agent is different, drawn from a given distribution 
\begin{equation} 
\Pi(\lambda)=(1+\delta) (1-\lambda)^{\delta}, {\rm with} -1< \delta < \infty,
\label{eq:lambdadist}
\end{equation}
where $\lambda$ is a fraction in the interval $[0,1-\epsilon]$, where 
$\epsilon$ is arbitrarily small and positive.
The dynamics of exchange follows
\begin{equation}
\begin{aligned}
m_i(t+1) &= \lambda_i m_i(t)+r((1-\lambda_i)m_i(t)+(1-\lambda_j)m_j(t)) \\
m_j(t+1) &= \lambda_j m_j(t)+(1-r)((1-\lambda_i)(m_i(t)+(1-\lambda_j)m_j(t)),
\end{aligned}
\label{eq:ccm}
\end{equation}
where $r$ is a random fraction in $[0,1]$, drawn in each time (exchange) step.
$\lambda_i$ is the saving fraction of agent $i$ whose wealth is $m_i(t)$
at trading step $t$. 
$\lambda_i$  are quenched and drawn randomly from $\Pi(\lambda)$ 
(Eq.~\ref{eq:lambdadist}).
The asymptotic form of the steady state distribution of wealth is given 
by~\cite{Chatterjee2007}
\begin{eqnarray}
P(m)\sim m^{-(2+\delta)}.
\label{eq:mdel}
\end{eqnarray}
\begin{figure}[t]
\begin{center}
\includegraphics[width=17.0cm]{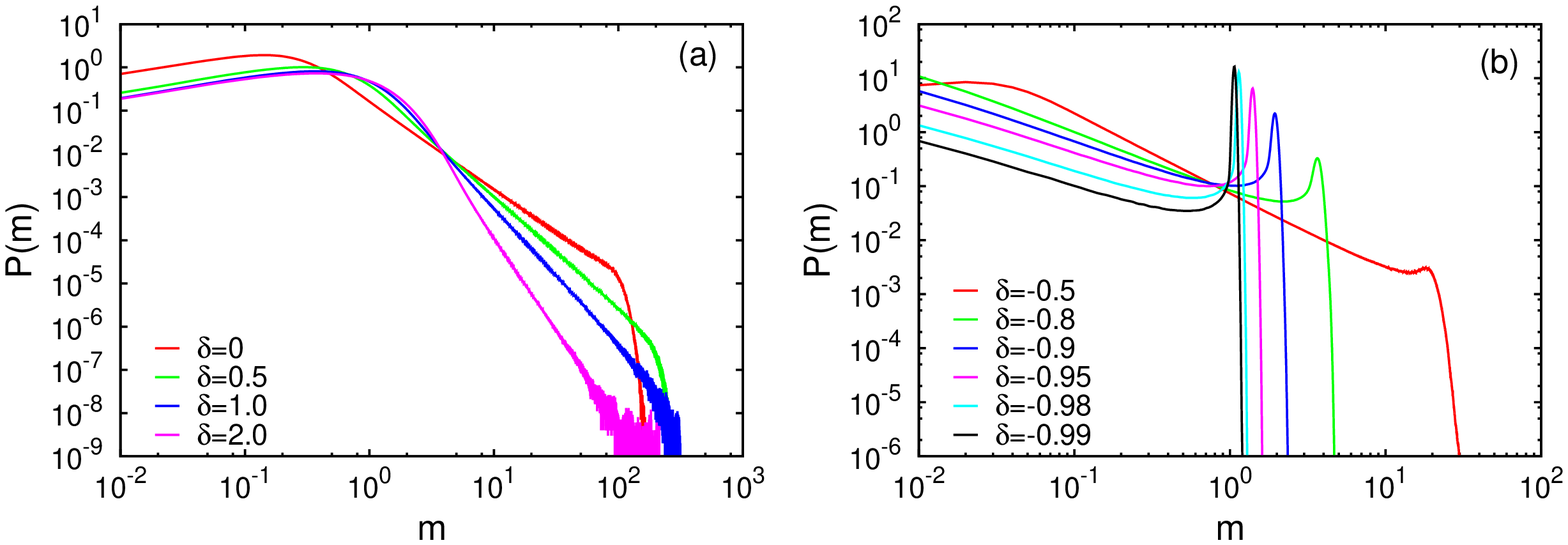}
\includegraphics[width=8.9cm]{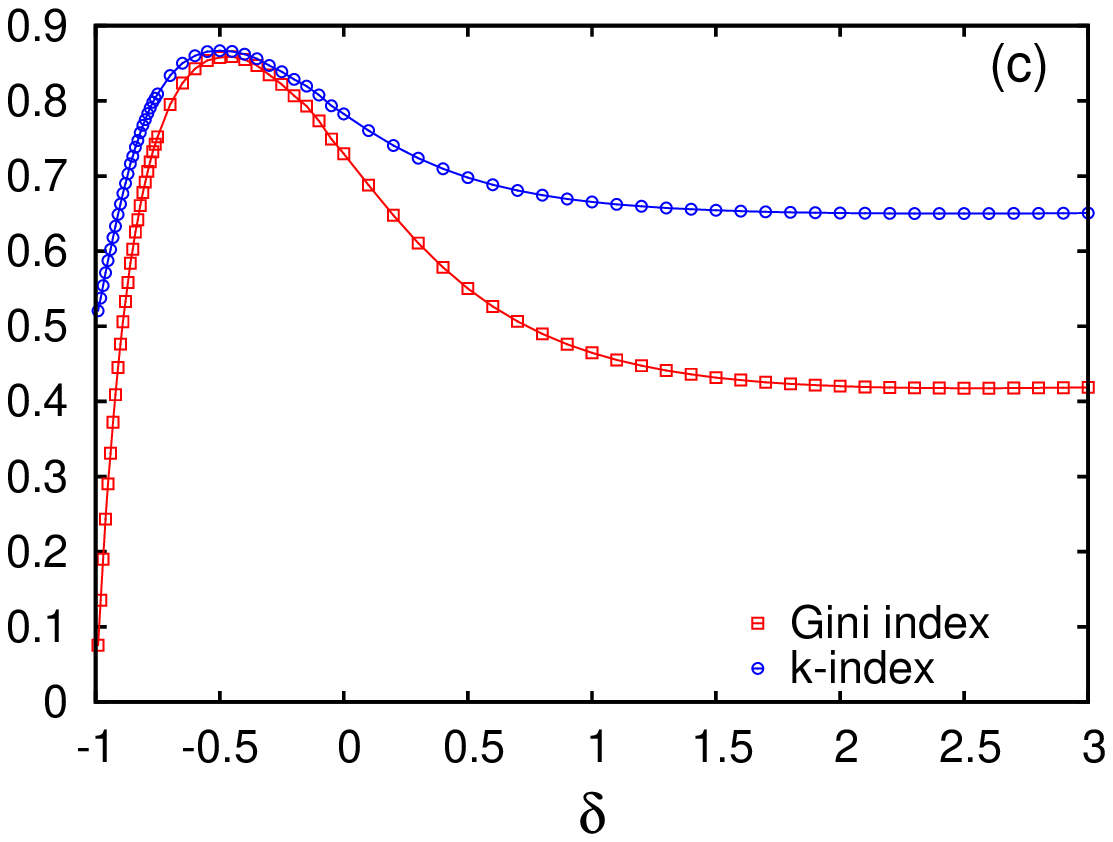}
\end{center}
\caption{ Probability distributions $P(m)$  in the CCM model for various 
distributions of $\lambda$, given by $\Pi(\lambda)=(1+\delta) 
(1-\lambda)^{\delta}$.
The power law exponents are $\nu=2+\delta$.
(a) For $\delta=0,0.5,1.0,2.0$; (b) For negative values of $\delta$.
(c) Gini and $k$-index for a range of $\delta$ values in the CCM model.
Simulation results are shown for $N=1000$ agents with average wealth unity.
}
\label{fig:ccm}
\end{figure}
We compute $P(m)$ vs. $m$ for different values of $\delta$.
In the simulations, we allow values of $\lambda$ until a certain fixed upper 
cutoff ($0.999$ for our case; i.e., $\epsilon=0.001$), because the population 
of agents with $\lambda$ close to $1$ is 
already high, and we have to restrict any agent assuming a saving propensity 
very close to unity. As a result, there is no finite size effect in the 
calculated values.

In Fig.~\ref{fig:ccm}a  we plot the distributions $P(m)$ for non-negative 
values of $\delta$. 
We find that the probability distribution $P(m)$ follows Eq.~\ref{eq:mdel} for 
most of the range of $m$, until an exponential cut-off, 
which is a result of the truncation of $\Pi(\lambda)$ close to $1$.
In Fig.~\ref{fig:ccm}b we plot the same for negative values of $\delta$. 
The probability distribution $P(m)$ again follows Eq.~\ref{eq:mdel} for most of 
the range of $m$. However, it is quite interesting to note that for $\delta 
\lesssim -0.5$, $P(m)$ shows a second peak at a large value of $m$, say $m_*$.
This $m_*$ moves towards $\langle m \rangle =1$ as $\delta \to -1$. This is 
quite easy to explain theoretically: as $\delta \to -1$, $\Pi(\lambda)$ is 
peaked near $\lambda \to 1$, essentially more and more fraction of agents have 
very high saving propensities, a situation similar to $\lambda \to 1$ in 
CC model. We recall that this situation will tend to produce $P(m)$ peaked at 
average money per agent ($\langle m \rangle =1$), which is equivalent to more 
``equality''.
$P(m)=\Delta(m-\langle m \rangle)$ with $\langle m \rangle=1$ for $\lambda=1$ 
in CC model; here $\Delta(\cdot)$ is the Dirac delta function. Then Gini index 
$g=0$ and $k=0.5$.
In comparison, $\lambda$ are distributed in CCM model and making $\delta \to -1$
makes $\Pi(\lambda)$ further peaked near $\lambda=1$, more agents have 
`similar' values of saving, close to unity and produce the second peak close to 
$m=\langle m \rangle=1$. The peak moves towards $\langle m \rangle$ as $\delta 
\to -1$. Additionally, Gini 
index $g \to 0$.
It may be noted that, because of this modification over the power 
law (Eq.~\ref{eq:mdel}) in the distribution function $P(m)$, the standard 
relationship between Gini index and the Pareto exponent (see e.g., 	
Ref.~\cite{chakrabarti2013econophysics}) is not valid here.

We measured inequality in the distributions in terms of Gini and $k$-indices 
and plotted in Fig.~\ref{fig:ccm}(c) for different values of $\delta$. 
Inequality 
seems to be maximum around $\delta \approx -0.5$

\subsection{Model for bimodal distribution and phase diagram}
\begin{figure}[h]
\includegraphics[height=7.5cm]{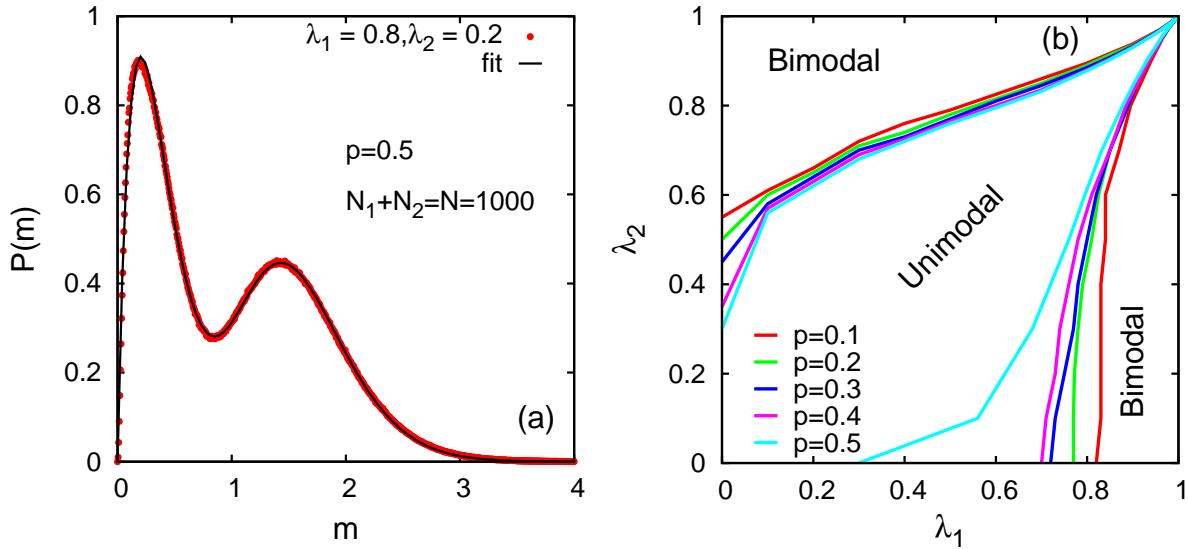}
\caption{(a)  Wealth distribution $P(m)$ for two groups of agents of same 
size, $N_1=N_2=500$, i.e., $p=0.5$ with saving propensities $\lambda_1=0.2$ 
and $\lambda_2=0.8$. The total system size is $N_1+N_2=N=10^3$. 
The steady state wealth distribution is fitted  to a combination of two 
Gamma distributions $a_{1} m^{\alpha_1} e^{-\beta_1 m} + a_{2} m^{\alpha_2} 
e^{-\beta_2 m} $ with parameter values $\alpha_1=0.98$, $\alpha_2=10.57$, 
$\beta_1=4.82$ and $\beta_2=7.23$.
(b) Phase boundaries for various values of $p$, showing regions of 
unimodal and bimodal distributions.
The simulations have been performed for $N=10^3$ agents.
}
\label{fig:bimodal}
\end{figure}
Bimodal distributions in wealth distributions are not 
uncommon~\cite{Wileybook2}, and is also observed in firm 
sizes~\cite{chakrabarti2013bimodality}. In the following we propose a very 
simple modification in the kinetic exchange model framework, to produce bimodal 
distribution of wealth.

\begin{figure}[h]
\includegraphics[width=17.9cm]{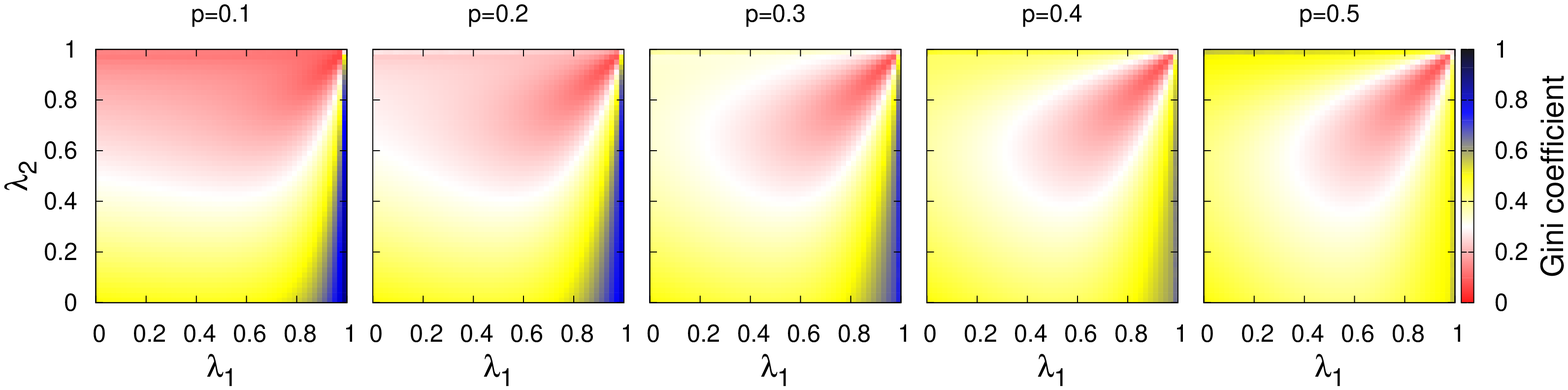}
\includegraphics[width=17.9cm]{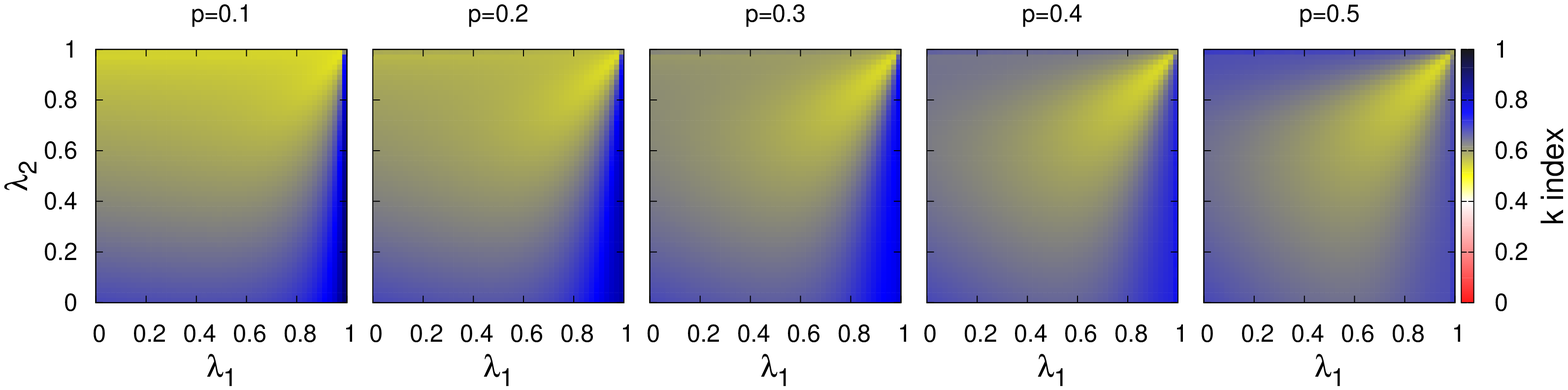}
\caption{Top: Gini indices for different cases 
 where $N_1$ agents have saving fraction $\lambda_1$ and $N_2$ 
agents have $\lambda_2$ each, for different values of $p=N_1/(N_1+N_2)$.
Bottom: $k$-index for the same. 
The simulations have been performed for $N_1+N_2=N=10^3$ agents.
}
\label{fig:bimodalgini}
\end{figure}

Let us now consider two  groups of  $N_1$ and $N_2$ agents, with saving 
propensities $\lambda_1$ and $\lambda_2$ respectively.
Let $p=N_1/(N_1+N_2)$. 
Agents' saving propensities remain unchanged over time, and they exchange money 
using the same rule as CC model (Eq.~\ref{eq:cc}). 
For example, let us consider the case $p=0.5$ i.e., $N_1=N_2$. Let 
$\lambda_1=0.2$ and $\lambda_2=0.8$. 
After exchanging their money, the steady state distribution is 
shown in Fig.~\ref{fig:bimodal}. The distribution clearly shows bimodal 
distribution. We fit the distribution with a combination of two Gamma 
distributions $a_{1} m^{\alpha_1} e^{-\beta_1 m} + a_{2} m^{\alpha_2} 
e^{-\beta_2 m} $ with parameter values $\alpha_1=0.98$, $\alpha_2=10.57$, 
$\beta_1=4.82$ and $\beta_2=7.23$  as shown in the Fig~\ref{fig:bimodal}a.
We found that all combinations of ($\lambda_1,\lambda_2$) do not give the 
bimodal distributions.
The combination of  values giving bimodal distribution are shown in the phase 
diagram. The boundary region is roughly estimated for 
various values of $p$ ($=0.1,0.2,0.3,0.4,0.5$) and shown in 
Fig.~\ref{fig:bimodal}b. 
We also computed Gini index and $k$-index for different combination of 
($\lambda_1,\lambda_2$) for various values of $p$ (Fig.~\ref{fig:bimodalgini}).

\section{Gamma distribution and its inequality statistics}
For the CC model, the steady state wealth distribution closely fits gamma 
distributions~\cite{Patriarca2004}. Let us compute the inequality measures 
considering such a distribution.
For the Gamma distribution: 
\begin{equation}
P(m) \propto m^{\alpha} {\rm e}^{-\beta m}, \,\,\,\beta \equiv 1/T
\label{eq:gamma}
\end{equation}
we evaluate the inequality statistics. 
The cumulative distribution is given as 
\begin{equation}
x( r) = 
\int_{0}^{r}m^{\alpha} {\rm e}^{-\beta m}dm =
\frac{\gamma (\alpha+1,\beta r)}{\beta^{1+\alpha}}
\end{equation}
where $\gamma (a,x)$ is an incomplete gamma function defined by 
\begin{equation}
\gamma (a, x) = \int_{0}^{x}
t^{a-1} {\rm e}^{-t}dt. 
\end{equation}
Hence, we have the normalization constant of the distribution (\ref{eq:gamma}) 
as 
\begin{equation}
x(\infty) = 
\frac{\Gamma (\alpha +1)}{\beta^{1+\alpha}}
\end{equation}
where we define the Gamma function by 
\begin{equation}
\Gamma (a) = \int_{0}^{\infty}t^{a-1}{\rm e}^{-t} dt = \gamma (a,\infty). 
\end{equation}
Thus, we have 
\begin{equation}
X(r ) \equiv \frac{x(r )}{x(\infty)} = 
\frac{\gamma(\alpha+1,\beta r)}{\Gamma (\alpha+1)}. 
\label{eq:def_X}
\end{equation}
Similarly, we have 
\begin{equation}
y(r )= 
\frac{1}{\beta^{2+\alpha}}
\int_{0}^{\beta r} m^{\alpha +1}{\rm e}^{-m}dm = 
\frac{\gamma (\alpha +2,\beta r)}{\beta^{2+\alpha}}
\end{equation}
and 
\begin{equation}
y(\infty) = 
\frac{\Gamma (\alpha +2)}{\beta^{2+\alpha}}
\end{equation}
Hence,
\begin{equation}
Y(r ) \equiv 
\frac{y(r )}{y (\infty)} = 
\frac{\gamma (\alpha +2,\beta r)}
{\Gamma (\alpha + 2)}. 
\label{eq:def_Y}
\end{equation}
\begin{figure}[t]
\includegraphics[width=15.0cm]{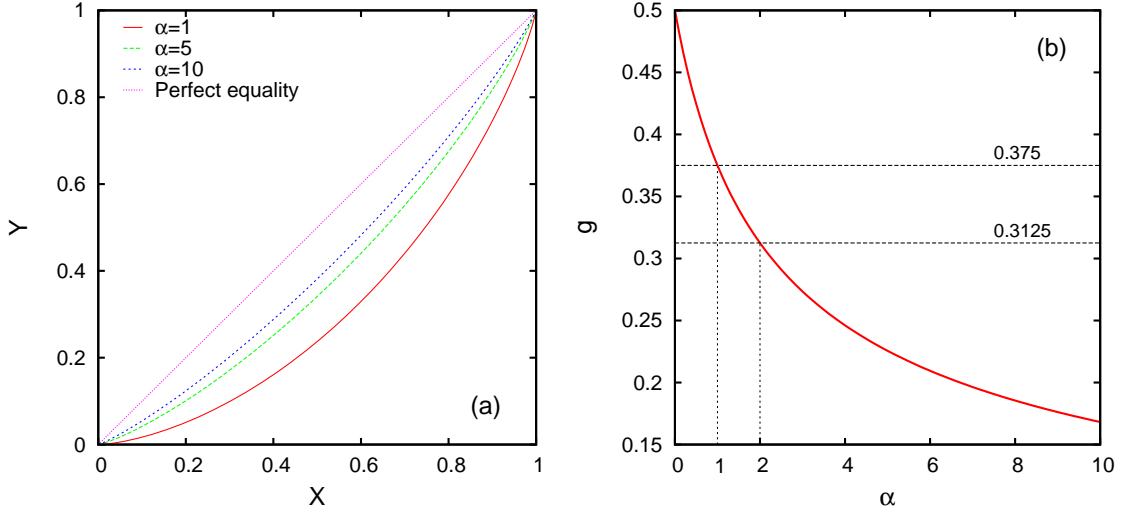}
\caption{Inequality in Gamma distributions: 
 (a) Lorentz curve for several values 
of $\alpha$; 
(b) Gini index $g$ as a function of $\alpha$.
}
\label{fig:fg5}
\end{figure}
Therefore, the Lorentz curve is given by 
(\ref{eq:def_X}) and (\ref{eq:def_Y}). 
In Fig.~\ref{fig:fg5}a, 
we plot the Lorentz curve for several values of 
$\alpha$. 
We also checked 
numerically that 
the curve is independent of the parameter $\beta$.
Next, to calculate Gini index, we check if the following relation is satisfied:
\begin{equation}
\gamma (a+1,x) = 
a\gamma (a,x) -x^{a}{\rm e}^{-x}
\end{equation}
Applying it to our case, we immediately obtain
\begin{equation}
\gamma (\alpha +2,\beta r) = 
(\alpha +1) 
\gamma (\alpha + 1,\beta r)-
(\beta r)^{\alpha +1}
{\rm e}^{-\beta r}. 
\end{equation}
This reads
\begin{equation}
\Gamma (\alpha +2) Y = 
(\alpha +1) \Gamma (\alpha +1) 
X -(\beta r)^{\alpha +1}{\rm e}^{-\beta r}. 
\end{equation}
If we notice $\Gamma (\alpha +2)=(\alpha +1)\Gamma (\alpha +1)$, 
we obtain
\begin{equation}
X-Y = 
\frac{(\beta r)^{\alpha +1} {\rm e}^{-\beta r}}
{\Gamma (\alpha +2)}. 
\end{equation}
Accompanying the derivative
\begin{equation}
\frac{dX}{dr}=\beta \frac{(\beta r)^{\alpha} {\rm e}^{-\beta r}}
{\Gamma (\alpha +1)}
\end{equation}
with $X-Y$, 
we get  the Gini index $g$ as
\begin{eqnarray}
g & = & 
2\int_{0}^{1}(X-Y)dX \nonumber \\
\mbox{} & = & 
2\beta  \int_{0}^{\infty}
\frac{(\beta r)^{\alpha +1} {\rm e}^{-\beta r}}
{\Gamma (\alpha +2)} 
\cdot 
\frac{(\beta r)^{\alpha} {\rm e}^{-\beta r}}
{\Gamma (\alpha +1)}dr \nonumber \\
\mbox{} & = & 
\frac{2\beta}{\Gamma (\alpha +2) \Gamma (\alpha +1)}
\int_{0}^{\infty}
(\beta r)^{2\alpha +1}
{\rm e}^{-2\beta r}dr =
\frac{\Gamma (2(\alpha +1))}
{2^{2\alpha +1} 
\Gamma (\alpha +1) \Gamma (\alpha +2)}. 
\end{eqnarray}
which is independent of $\beta$ and we 
recover the exponential case by setting $\alpha=0$ as 
\begin{equation}
g_{0} = 
\frac{\Gamma (2)}
{2 \Gamma (1) \Gamma (2)}=\frac{1}{2}.
\end{equation}
Note $\Gamma (1)=1$. 
The cases of $\alpha=1,2$ are given by 
\begin{eqnarray}
g_{1} & = & 
\frac{\Gamma (4)}{2^{3} \Gamma(2) \Gamma(3)}=
\frac{3}{8} \\
g_{2} & = & 
\frac{\gamma (6)}{2^{5} \Gamma (3) \Gamma (4)} = \frac{5}{16}
\end{eqnarray}
where we used 
$\Gamma (\alpha+1)=\alpha \Gamma (\alpha)$ recursively. 
In Fig.~\ref{fig:fg5}, 
we plot the $g$ as a function of $\alpha$, and values of $g$ corresponding to 
$\alpha=1,2$ are also indicated.

\section{Mixture of Gamma distributions: Unimodal and bimodal distribution}
\begin{figure}[hb]
\includegraphics[width=15.0cm]{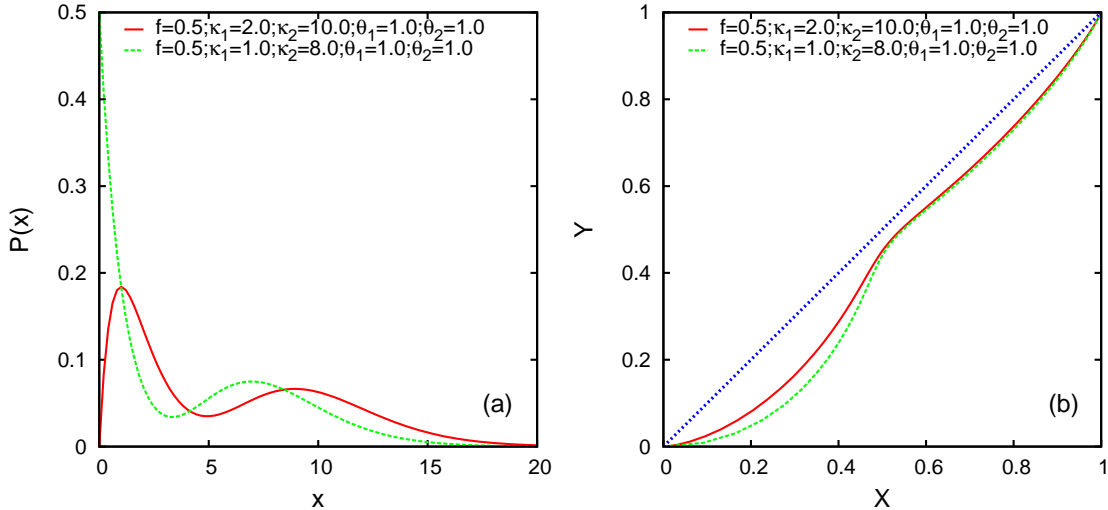}
\caption{(a)
Mixture of Gamma distributions  as a bimodal distribution 
(Eq.~\ref{eq;gammamix}). 
(b) Lorentz curve for the same (from Eq.~\ref{eq:xrbimod} and \ref{eq:yrbimod}).
}
\label{fig:fg6}
\end{figure}
We next consider the mixture of Gamma distribution as 
\begin{equation}
P(x) = 
(1-f) x^{\kappa_1-1}\frac{{\rm e}^{-x/\theta_{1}}}
{\Gamma (\kappa_1) \theta_{1}^{\kappa_1}} + 
f x^{\kappa_2-1}\frac{{\rm e}^{-x/\theta_{2}}}
{\Gamma (\kappa_2) \theta_{2}^{\kappa_2}} 
\label{eq;gammamix}
\end{equation}
Obviously, 
for $f=0$, the unimodal Gamma distribution takes its maximum at
$x = \theta_{1} (\kappa_1-1)$.
However, for $a\neq 1$, the peaks are located at the 
$x$ as solutions of 
\begin{equation}
\frac{1-f}
{\Gamma (\kappa_1) \theta_{1}^{\kappa_1+1}}
\left(\theta_{1}(\kappa_1-1)-x \right)x^{\kappa_1-1} {\rm e}^{-x/\theta_{1}}
+
\frac{f}{\Gamma (\kappa_2)\theta_{2}^{\kappa_2+1}}
\left(\theta_{2}(\kappa_2-1)-x\right)x^{\kappa_2-1}{\rm e}^{-x/\theta_{2}} =0 
\end{equation}
Hence, the locations of the peaks 
are dependent on the choice of parameters 
$\theta_{1},\theta_{2},\kappa_1,\kappa_2,f$. 
For the Lorentz curve, we obtain 
\begin{eqnarray}
X(r) & = & 
\frac{(1-f)}{\Gamma (\kappa_1)}
\gamma (\kappa_1,r/\theta_{1})
+
\frac{f}{\Gamma (\kappa_2)}
\gamma (\kappa_2,r/\theta_{2}) \label{eq:xrbimod}\\
Y (r ) & = & 
\frac{
\frac{(1-f)}{\Gamma (\kappa_1)}
\theta_{1} 
\gamma (\kappa_1+1, r/\theta_{1})
+
\frac{f}{\Gamma (\kappa_2)}
\theta_{2} \gamma (\kappa_2+1,r/\theta_{2})}
{Y (\infty)}
\label{eq:yrbimod}
\end{eqnarray}
where we defined 
\begin{equation}
Y(\infty) = 
(1-f)\theta_{1}
\left\{
\frac{\Gamma (\kappa_1+1)}
{\Gamma (\kappa_1)}
\right\}
+
f\theta_{2} 
\left\{
\frac{\Gamma (\kappa_2+1)}{\Gamma (\kappa_2)}
\right\}. 
\end{equation}
In Fig~\ref{fig:fg6}b, 
we plot the Lorentz curve for several choices of the parameters. 
For the above Lorentz curve, 
the Gini index $g(f,\kappa_1,\kappa_2,\theta_1,\theta_2)$ is calculated as 
\begin{eqnarray}
g & = & 
\left\{
\frac{1-f}{\Gamma (\kappa_1)}
\right\}^{2}
\frac{1}{\theta_{1}^{\kappa_1}}
\int_{0}^{\infty}
dr \, 
r^{\kappa_1-1}
{\rm e}^{-r/\theta_{1}}
\{
\gamma (\kappa_1,r/\theta_{1})
-\frac{\theta_{1}}{Y(\infty)}  \gamma (\kappa_1+1,r/\theta_{1})
\}  \nonumber \\
\mbox{} & + & 
\frac{f (1-f)}
{\theta_{1}^{\kappa_1}
\Gamma (\kappa_1)\Gamma (\kappa_2)}
\int_{0}^{\infty}
dr \, 
r^{\kappa_1-1}
{\rm e}^{-r/\theta_{1}}
\{
\gamma (\kappa_2,r/\theta_{2})
-\frac{\theta_{2}}{Y(\infty)}  \gamma (\kappa_2+1,r/\theta_{2})
\}    \nonumber \\
\mbox{} & + & 
\frac{f (1-f)}
{\theta_{2}^{\kappa_2}
\Gamma (\kappa_1)\Gamma (\kappa_2)}
\int_{0}^{\infty}
dr \, 
r^{\kappa_2-1}
{\rm e}^{-r/\theta_{2}}
\{
\gamma (\kappa_1,r/\theta_{1})
-\frac{\theta_{1}}{Y(\infty)} \gamma (\kappa_1+1,r/\theta_{1})
\}  \nonumber \\
\mbox{} & + & 
\left\{
\frac{f}{\Gamma (\kappa_2)}
\right\}^{2}
\frac{1}{\theta_{2}^{\kappa_2}}
\int_{0}^{\infty}
dr \, 
r^{\kappa_2-1}
{\rm e}^{-r/\theta_{2}}
\{
\gamma (\kappa_2,r/\theta_{2})
-\frac{\theta_{2}}{Y(\infty)} \gamma (\kappa_2+1,r/\theta_{2})
\}.
\end{eqnarray}
Knowing the parameters $f,\kappa_1,\kappa_2,\theta_1,\theta_2$, one can 
compute $g$ numerically from the above expression.

\section{Discussion}
Empirical data~\cite{worldbank} shows that Gini index varies mostly in 
$0.2-0.7$ 
(see Fig.~\ref{fig:emp}b).
The CC model gives Gini index in the range $0-0.5$. The inequality decreases 
monotonically with increasing 
saving propensity $\lambda$. $g=0.5$ for $\lambda=0$, the wealth distribution 
$P(m)$ is a perfect exponential distribution, giving the maximum value of 
inequality for this model. 
On the other extreme, when $\lambda \to 1$, $P(m)$ approaches a 
Dirac $\delta$-function $\Delta(m-\langle m \rangle)$, for which $g \to 0$ 
(Fig.~\ref{fig:cc}).
Hence this model does not  reproduce most of the range of real Gini indices.
In reality, Gini index rarely go below $0.3$, but often goes beyond $0.5$.

In CCM model, however, the range of the Gini index is quite wide, and in fact, 
overlaps
with almost the entire range of empirically observed Gini index values.
In fact, in the asymptotic limit of Eq.~\ref{eq:mdel} for practically infinite 
value of $\delta$,
$P(m)$ should approach an uniform distribution in $[0,1]$, which will yield a 
value of Gini index equal to $1/3$.
In Fig.~\ref{fig:ccm}b, we observe that for large values of 
$\delta$,  there is a tendency to saturate to a value close to $0.4$, which we 
anticipate, might as well approach $1/3$ for $\delta \to \infty$.

In some empirical analysis, the bulk of the wealth distribution resembles Gamma 
distribution. We analytically computed the Lorenz curve and the Gini index for 
Gamma distributions. There are even some instances where the wealth 
distribution are found to be double peaked~\cite{Wileybook2}. We propose a 
variation of the kinetic exchange models to model this, and a combination of 
Gamma distributions to fit the resulting distribution. The steady state 
distribution is unimodal or bimodal depending on the combination of values of 
the saving propensities and the relative fraction of agents of the two groups.
The phase boundaries for specific cases have been computed using numerical 
simulations.
We also show that it is possible to derive an exact expression for the Gini 
index, considering Gamma distribution as the best fit to the numerically 
computed steady state wealth distributions.

The critical studies of kinetic exchange models of wealth distributions seem to 
yield more and more interesting aspects, not only in terms of theoretical 
understanding of the models, but also when compared to empirical data.
For instance, one of the recent studies explain city size statistics using the 
same framework~\cite{ghosh2014zipf}. 
There has not been only a very few studies~\cite{chakrabarti2010inequality} 
that discuss inequality measures in reference to models.
Further research will be able to elucidate 
the usefulness of such a simple framework in understanding complex 
socio-economic phenomena.

\begin{acknowledgments}
A.C. and B.K.C. acknowledge support from B.K.C.'s J. C. Bose Fellowship
Research Grant.
\end{acknowledgments}

%

\begin{thebibliography}{10}
\expandafter\ifx\csname url\endcsname\relax
  \def\url#1{\texttt{#1}}\fi
\expandafter\ifx\csname urlprefix\endcsname\relax\def\urlprefix{URL }\fi

\bibitem{arrow2000meritocracy}
K.~J. Arrow, S.~Bowles, S.~N. Durlauf, Meritocracy and economic inequality,
  Princeton Univ. Press, 2000.

\bibitem{stiglitz2012price}
J.~E. Stiglitz, The price of inequality: How today's divided society endangers
  our future, WW Norton \& Company, 2012.

\bibitem{neckerman2004social}
K.~Neckerman, Social Inequality, Russell Sage Foundation, 2004.

\bibitem{goldthorpe2010analysing}
J.~H. Goldthorpe, Analysing social inequality: a critique of two recent
  contributions from economics and epidemiology, Eur. Sociological Rev. 26~(6)
  (2010) 731--744.

\bibitem{hurst1995social}
C.~E. Hurst, Social Inequality: Forms, Causes, and Consequences, Allyn and
  Bacon, Boston, 1995.

\bibitem{Yakovenko:RMP}
V.~M. Yakovenko, J.~Barkley Rosser~Jr., Colloquium: Statistical mechanics of
  money, wealth and income, Rev. Mod. Phys. 81 (2009) 1703--1725.

\bibitem{chakrabarti2013econophysics}
B.~K. Chakrabarti, A.~Chakraborti, S.~R. Chakravarty, A.~Chatterjee,
  Econophysics of income and wealth distributions, Cambridge Univ. Press,
  Cambridge, 2013.

\bibitem{aoyama2010econophysics}
H.~Aoyama, Y.~Fujiwara, Y.~Ikeda, Econophysics and companies: statistical life
  and death in complex business networks, Cambridge Univ. Press, Cambridge,
  2010.

\bibitem{chatterjee2014socio}
A.~Chatterjee, Socio-economic inequalities: a statistical physics perspective,
  in: Econophysics and Data Driven Modelling of Market Dynamics, Eds. F
  Abergel, H. Aoyama, B.K. Chakrabarti, A. Chakraborti, A. Ghosh,, New Economic
  Windows, Springer (2015), 2014.

\bibitem{chatterjee2015social}
A.~Chatterjee, A.~Ghosh, J.-I. Inoue, B.~K. Chakrabarti, Social inequality:
  from data to statistical physics modeling, J. Phys. Conf. Ser. 638 (2015)
  012014.

\bibitem{Cho23052014}
A.~Cho, Physicists say it's simple, Science 344~(6186) (2014) 828.
\newline\urlprefix\url{http://www.sciencemag.org/content/344/6186/828.short}

\bibitem{Chin23052014}
G.~Chin, E.~Culotta, What the numbers tell us, Science 344~(6186) (2014)
  818--821.
\newline\urlprefix\url{http://www.sciencemag.org/content/344/6186/818.short}

\bibitem{Xie23052014}
Y.~Xie, Undemocracy: Inequalities in science, Science 344~(6186) (2014)
  809--810.
\newline\urlprefix\url{http://www.sciencemag.org/content/344/6186/809.short}

\bibitem{Pareto-book}
V.~Pareto, Cours d'economie politique, Rouge, Lausanne, 1897.

\bibitem{Montroll:1982}
E.~W. Montroll, M.~F. Shlesinger, On 1/f noise and other distributions with
  long tails, Proc. Natl. Acad. Sci. 79 (1982) 3380--3383.

\bibitem{gini1921measurement}
C.~Gini, Measurement of inequality of incomes, Econ. J. 31~(121) (1921)
  124--126.

\bibitem{Hogg-2007}
R.~Hogg, J.~Mckean, A.~Craig, {Introduction to mathematical statistics},
  Pearson Education, Delhi, 2007.

\bibitem{Chatterjee:EWD}
A.~Chatterjee, S.~Yarlagadda, B.~K. Chakrabarti (Eds.), Econophysics of Wealth
  Distributions, New Economic Windows Series, Springer-Verlag, Milan, 2005.

\bibitem{Chatterjee2007}
A.~Chatterjee, B.~K. Chakrabarti, Kinetic exchange models for income and wealth
  distributions, Eur. Phys. J. B 60~(2) (2007) 135--149.

\bibitem{Wileybook2}
P.~Richmond, S.~Hutzler, R.~Coelho, P.~Repetowicz, A review of empirical
  studies and models of income distributions in society, in: B.~K. Chakrabarti,
  A.~Chakraborti, A.~Chatterjee (Eds.), Econophysics and Sociophysics: Trends
  and Perspectives, Wiley-VCH, Weinheim, 2007, pp. 131--159.

\bibitem{Dragulescu:2000}
A.~A. Dr\u{a}gulescu, V.~M. Yakovenko, Statistical mechanics of money, Eur.
  Phys. J. B 17 (2000) 723--729.

\bibitem{Chakraborti:2000}
A.~Chakraborti, B.~K. Chakrabarti, Statistical mechanics of money: how saving
  propensity affects its distribution, Eur. Phys. J. B 17 (2000) 167--170.

\bibitem{Patriarca2004}
M.~Patriarca, A.~Chakraborti, K.~Kaski, {Statistical model with a standard
  $\Gamma$ distribution}, Phys. Rev. E 70~(1) (2004) 016104.

\bibitem{Chatterjee:2004}
A.~Chatterjee, B.~K. Chakrabarti, S.~S. Manna, Pareto law in a kinetic model of
  market with random saving propensity, Physica A 335 (2004) 155--163.

\bibitem{pareschi2013interacting}
L.~Pareschi, G.~Toscani, Interacting Multiagent Systems: Kinetic Equations and
  Monte Carlo Methods, Oxford Univ. Press, Oxford, 2013.

\bibitem{theil1967economics}
H.~Theil, Economics and information theory, North-Holland Amsterdam, 1967.

\bibitem{eliazar2010measuring}
I.~I. Eliazar, I.~M. Sokolov, Measuring statistical heterogeneity: The pietra
  index, Physica A 389~(1) (2010) 117--125.

\bibitem{ghosh2014inequality}
A.~Ghosh, N.~Chattopadhyay, B.~K. Chakrabarti, Inequality in societies,
  academic institutions and science journals: Gini and k-indices, Physica A
  410~(14) (2014) 30--34.

\bibitem{druagulescu2001exponential}
A.~A. Dr{\u{a}}gulescu, V.~M. Yakovenko, Exponential and power-law probability
  distributions of wealth and income in the united kingdom and the united
  states, Physica A 299~(1) (2001) 213--221.

\bibitem{Lorenz}
M.~O. Lorenz, Methods for measuring the concentration of wealth, Am. Stat.
  Assoc. 9 (1905) 209--219.

\bibitem{inoue2015measuring}
J.-I. Inoue, A.~Ghosh, A.~Chatterjee, B.~K. Chakrabarti, Measuring social
  inequality with quantitative methodology: analytical estimates and empirical
  data analysis by gini and $ k $ indices, Physica A 429 (2015) 184--204.

\bibitem{worldbank}
{World Bank, All the Ginis Dataset, retrieved June, 2014},
  
http://siteresources.worldbank.org/INTRES/Resources/469232-1107449512766/allgini
s{\_}2013.xls.

\bibitem{chakrabarti2013bimodality}
A.~S. Chakrabarti, Bimodality in the firm size distributions: a kinetic
  exchange model approach, Eur. Phys. J. B 86~(6) (2013) 1--6.

\bibitem{ghosh2014zipf}
A.~Ghosh, A.~Chatterjee, A.~S. Chakrabarti, B.~K. Chakrabarti, Zipf's law in
  city size from a resource utilization model, Phys. Rev. E 90~(4) (2014)
  042815.

\bibitem{chakrabarti2010inequality}
A.~S. Chakrabarti, B.~K. Chakrabarti, Inequality reversal: Effects of the
  savings propensity and correlated returns, Physica A 389~(17) (2010)
  3572--3579.

\end{thebibliography}

%

\end{document}